\documentstyle[12pt,epsfig]{article}
\begin{document}
\title{Broad Histogram Method for Continuous Systems : The XY-Model}

\author{Jos\'e~D.~Mu\~noz$^{1,2}$, Hans~J.~Herrmann$^{1,3}$\\
  \small (1) Institute for Computer Applications 1, Stuttgart University,\\
  \small Pfaffenwaldring 27, D-70569 Stuttgart, Germany\\
  \small (2) Permanent address: Dpto. de F\'{\i}sica, Universidad Nacional de
  Colombia,\\ 
  \small Bogot\'a D.C., Colombia\\
  \small E-mail: jdmunoz@ica1.uni-stuttgart.de\\
  \small  (3) E-mail hans@ica1.uni-stuttgart.de}
\date{September 22th, 1998}
\maketitle

\begin{abstract}
We propose a way of implementing the Broad Histogram Monte Carlo method to
systems with continuous degrees of freedom, and we apply these ideas to
investigate the three-dimensional XY-model with periodic boundary
conditions. We have found an excellent agreement between our method
and traditional Metropolis results for the energy, the
magnetization, the specific heat and the magnetic
susceptibility on a very large temperature range. For the calculation
of these quantities in the temperature range $0.7<T<4.7$ our
method took less CPU time than the Metropolis simulations for $16$
temperature points in that temperature range. Furthermore,
it calculates the whole temperature range
$1.2<T<4.7$ using only $2.2$ times more computer effort than the
Histogram Monte Carlo method for the range $2.1<T<2.2$. 
Our way of
treatment is general, it can also be applied to other 
systems with continuous degrees of freedom.
\end{abstract}

\newpage
\section{Introduction}

For thermodynamic systems in equilibrium one
typically wants to calculate average values of quantities $Q$ like the
energy, the magnetization, etc.. 
These mean values depend on one hand of the  properties of the system
himself and on the other hand on its interactions with the environment. 
For example, in the case of the canonical ensemble, the system is in
equilibrium with a heat bath at temperature $T$, and the mean value
$<Q>_T$ is given by 
\begin{equation}
  \label{canonica1}
  <Q>_T={\sum_E <Q>_E g(E) \exp{(-E/k_BT)}  
         \over 
         \sum_Eg(E) \exp{(-E/k_BT)}},
\end{equation}
where the degeneracy $g(E)$ is the number of distinct states with
energy $E$, $<Q>_E$ denotes 
the micro-canonical average of $Q$ at energy $E$ and $k_B$ is the
Boltzmann constant (in the rest of this paper $k_B=1$ is taken). Here,
$g(E)$ and $<Q>_E$ are intrinsic characteristics of the system, and
$\exp{(-E/k_BT)}$ is the Boltzmann factor.

Most Monte Carlo methods were designed to calculate particular
averages like Eq. (\ref{canonica1}).
For example, Metropolis algorithms \cite{Metropolis} and similar
canonical simulations accumulate sampling distributions that
approximate Eq. (\ref{canonica1}) for a large number of samples and,
therefore, the mean values $<Q>_T$ are estimated simply by averaging
$Q$ on the sample set. However, if one wants to study the dependence of
$<Q>_T$ on $T$ using this kind of algorithms it is necessary to run
one entire simulation at each temperature value, with large computer
efforts.

A well-proved strategy to avoid these multiple calculations is the
histogram method (HMC), introduced by Salsburg \cite{his1} and popularized by
Ferrenberg and Swendsen \cite{his2}.
This method reweights data of one canonical simulation at one temperature
$T_{\rm o}$ to find mean values at other temperatures. 
However, a canonical distribution has a very narrow peak on the energy
axis around its mean energy and, therefore, there are not enough
samples in the tails of the distribution to have good statistics
there.
This restricts the reliability of the method to a relatively
narrow temperature range around $T_{\rm o}$ \cite{hisprobl,HMCerror}. 

The Broad Histogram Monte Carlo Method (BHMC) \cite{bhmc1}-\cite{BHMCisExact}  was
established by de Oliveira {\it et. al.} in order to overcome these
restrictions. 
It is designed to determine the system density of states $g(E)$ and
$<Q>_E$ directly.
These quantities are calculated from micro-canonical averages and,
therefore, samples with different energies can be considered
independently. The samples can be obtained
in many different ways, for
instance, performing micro-canonical simulations \cite{bhmc3} or
performing a non-biased random walk along the energy 
axis, which gives rise to much broader histograms than the usual
histogram method \cite{bhmc1,bhmc2}.
Given $g(E)$ and $<Q>_E$, the canonical distribution and the desired
mean values for any temperature can be obtained. Thus, it is possible 
to calculate mean values over the whole temperature range of interest
in only one run. The BHMC method was first developed for
systems with discrete degrees of freedom, and has shown to give results
more accurately and using less computer time than the conventional
histogram method for the 2d an 3d Ising model \cite{bhmc1}-\cite{bhmc3}, the
Edwards Anderson spin glass \cite{bhmc2} and the 3d Potts glass
\cite{potts}.

In the present paper we propose a way of extending the method to
systems with continuous degrees of freedom, using the XY-model as
testing ground.
Our way of treatment is general, it can also be applied to other
systems of this kind.
The remaining part of the paper is organized as follows.
In Sec. \ref{SecBHMC} we briefly summarize the basis of the BHMC
method.
In Sec. \ref{SecXY} we describe the XY-model and we use it as
example to show how to apply the method to continuous systems.
In Sec. \ref{Sampling} we discuss several ways of taking the
samples. 
In Sec. \ref{SecImplementa} we give technical details
about the implementation we did of the method. 
In Sec. \ref{results} we present the results obtained
with the BHMC method from a 3D XY-model on a $10 \times 10 \times 10$
cubic lattice 
with periodic boundary conditions, and we compare these results with
those of Metropolis simulations and of the histogram method.
On one side, we have obtained an excellent agreement between the BHMC
method and Metropolis simulations in all the calculated quantities
over a very broad temperature range,
i.e. for temperatures between $0.7$ and $4.7$, using
less computer time for the BHMC method than the time required for $16$
points by Metropolis simulations. On the other side, it reproduces
with the same accuracy the results of the HMC method on the
temperature range $2.1<T<2.2$ but, instead of diverging for temperatures
out of this range as the HMC method does, it remains precise over the
whole range $1.2<T<4.7$ using only $2.2$ times more computer effort
than the HMC method. It justifies the name ``Broad Histogram'' for the
method.
Finally, in Sec. \ref{conclusions} we summarize the main steps
of the proposed strategy to apply the BHMC method to continuous
systems and we discuss other strategies that can be used to implement
the method.

\section{The BHMC Method}
  \label{SecBHMC}

The strategy of the BHMC method is to calculate the degeneracy of
energy states $g(E)$ directly. The method for systems with discrete
degrees of freedom can be summarized as follows.

First, let us imagine a protocol of allowed movements in the space of
states of the system such that changing from an $X_{old}$ to an
$X_{new}$ configuration is allowed if and only if the reverse change
is also allowed, i.e.
\begin{eqnarray}
  \label{microreversible}
  X_{old} \to X_{new} \Longleftrightarrow X_{new} \to X_{old} \\
  \mbox{is allowed} \qquad \quad \mbox{is allowed} \nonumber
\end{eqnarray}
In other words, the protocol 
is micro-reversible. These movements are only virtual, in the sense
that they are not performed. They are introduced only to estimate the
density of states g(E).

Next, let us choose a fixed amount of energy change $\Delta E_{fix}$
and compute $N_{up}(X)$ ($N_{dn}(X)$) for the configuration X as
the number of allowed changes that increases (decreases) the energy of
the configuration by $\Delta E_{fix}$. Let ${<N_{up}(E)>}$
($<N_{dn}(E)>$) be the micro-canonical average of $N_{up}(X)$
($N_{dn}(X)$) at energy $E$. 

Due to Eq. (\ref{microreversible}),
the total number of ways to go up, summed over all states with energy
$E$, is equal to the total number of ways to go down, summed over all
states with energy $E+\Delta E_{fix}$, i.e. 
\begin{equation}
 \label{basic}
   g(E) <N_{up}(E)> = g(E+\Delta E_{fix})
   <N_{dn}(E+\Delta E_{fix})>.
\end{equation}
Eq. (\ref{basic}) can be rewritten as:
\begin{equation}
  \label{diff}
 \ln g(E+\Delta E_{fix})-\ln g(E) = 
 \ln {<N_{up}(E)> \over
   <N_{dn}(E+\Delta E_{fix})>}. 
\end{equation}
Hence, knowing $<N_{up}(E)>$ and $<N_{dn}(E)>$ allows to calculate
the right hand side of Eq. (\ref{diff}) and obtains $\ln g(E)$ for all
values of $E$ in steps of $\Delta E_{fix}$ by
adding up differences from a given initial point $(E_{\rm o},\ln
g(E_{\rm o}))$.

Similarly, it can be observed that Eq. (\ref{diff}) divided by
$\Delta E_{fix}$ estimates $\beta(E)$, i.e.
\begin{eqnarray}
 \label{beta}
 \beta(E) \equiv {d \ln g(E) \over dE} &\simeq&
 {\ln g(E+\Delta E_{fix})-\ln g(E) \over   
   \Delta E_{fix}} \nonumber \\ &\simeq&
 {1 \over \Delta E_{fix}} \ln {<N_{up}(E)>
   \over
   <N_{dn}(E+\Delta E_{fix})>}.
\end{eqnarray}
Then, $\beta (E)$ can be obtained for a set of energy values
$\{E_1 , E_2 , \cdots , E_n \}$ by estimating $<N_{up}(E_i)>$
and $<N_{dn}(E+\Delta {E_i}_{fix})>$ for each value
$E_i$. Eq. (\ref{diff}) can be used to estimate $\beta(E)$ in more
accurate ways, as it will be discussed in Sec. \ref{Sampling}. 
$\ln g(E)$ can be obtained, thereafter, by integrating $\beta (E)$
numerically from a given initial point $(E_{\rm o},\ln g(E_{\rm o}))$.

The value of $\ln g(E_{\rm o})$ for the initial point does not
affect the calculation of mean values $<Q>_T$, as can be shown in
Eq. (\ref{canonica1}). 
If $\ln g(E_{\rm o})=0$ is taken, the method gives $g(E)/g(E_{\rm
  o})$ for each energy.

Finally, to carry out calculations with this method  four
histograms are required, i.e. $N_{up}(E)$, $N_{dn}(E)$, $Q(E)$ and
the number of visits $V(E)$.  
Estimations for $<N_{up}(E)>$ and $<N_{dn}(E)>$ and, in consequence,
an estimation of $g(E)$ are obtained by dividing $N_{up}(E)$ and
$N_{dn}(E)$ by $V(E)$. 
$<Q>_E$ is calculated by dividing $Q(E)$ by $V(E)$.
Combining these two results Eq. (\ref{canonica1}) can be used to
calculate $<Q>_T$ at any desired temperature. 

It can be shown that Eq. (\ref{basic}) is exact for any statistical
model \cite{BHMCisExact}. Monte Carlo sampling processes are used only
to estimate the micro-canonical averages $<N_{up}(E)>$ and
$<N_{dn}(E)>$. Many different sampling strategies can be employed, as
will be discussed in Secs. \ref{Sampling} and \ref{conclusions}. 
Nevertheless, all states with the same energy have to be sampled with
the same probability and the correlations between samples have to be
small enough to obtain good estimations of such micro-canonical
averages. Otherwise, problems can appear \cite{RWproblem,BHMCisExact}.

\section{The BHMC method and the XY-Model}
  \label{SecXY}

\subsection{The XY-model}

In order to describe our technique for extending the BHMC method to
systems with continuous degrees of freedom we have chosen the XY-model
as testing ground.
The XY-model \cite{XY_Li_godM}-\cite{XY_Adler}, consists of a
set of spins $\vec{\sigma}$ of length unity arranged on a lattice. 
Each spin is allowed to rotate in a plane, 
characterized by an angle $\theta$. These variables define the state
of the system and can change continuously in the range $\theta \in
[-\pi,\pi]$. The Hamiltonian with zero external field is given by 
\begin{equation}
  \label{Hamiltonian}
  \mathcal{H}= -{\it J} \sum_{<{\it i j}>} \vec{\sigma}_{\it i} \cdot
  \vec{\sigma}_{\it j} 
  = -{\it J} \sum_{ < {\it  ij} > } \cos (\theta_{\it i} - \theta_{\it j}) 
\end{equation}
where the summations $<ij>$ are taken over all pairs of
nearest-neighbor sites, $\vec{\sigma}_i \cdot \vec{\sigma}_j$ is the
scalar product between $\vec{\sigma}_i$ and $\vec{\sigma}_j$, and $J$
denotes the maximal energy per bond (in the rest of this paper we will
take $J=1$). 
The order parameter chosen for this model \cite{XY_Li_godM} is the
average magnetization $M$, defined by:
\begin{equation}
  M \equiv \sum_i A_i \quad ,
\end{equation}
with
\begin{equation}
 \label{A}
   A_i \equiv \sqrt{\left( \sum_{j} \cos \theta_j
   \right)^2 + \left( \sum_{j} \sin \theta_j \right)^2}
\end{equation}
where the summations $j$ are taken over all nearest neighbors of the
site $i$. The XY-model so defined is one of the simplest
thermodynamic systems with continuous degrees of freedom.

\subsection{The BHMC method for continuous systems}
\label{BHMCcontinuo}

In order to extend the BHMC method for continuous systems, it is necessary to
redefine the quantities involved in Eq. (\ref{basic}) and to find a
condition for the protocol of movements equivalent to
Eq. (\ref{microreversible}) such that Eq. (\ref{basic}) holds.

First, let us imagine a protocol of random movements in the space
of states of the system, such that for each allowed movement the
probability to perform it equals the probability to revert it.
These movements are again only virtual, in the sense that they are
not executed. They are introduced only to estimate the density
of states g(E).
For the XY-model, for example, we use the following protocol:  
we choose one spin
at random with angle $\theta_{old}$, and then we choose for it a
new angle $\theta_{new}$ according to a uniform probability
distribution over the  range $\theta_{new} \in [-\pi ,+\pi]$.
It has to be noted that $\theta_{new}$ is a
random variable, with a probability density function (p.d.f) $f_{
\theta_{new}} (\theta)$ defined as the probability of obtaining
a value $\theta_{new}$ between $\theta$ and $\theta +d \theta$
given by 
\begin{equation}
  \label{fThetaNew}
  f_{\theta_{new}} (\theta)=
  \left\{
    \begin{array}{ccc}
      1/2 \pi & : & -\pi<\theta_{new}<\pi
      \\ 0    & : & otherwise
    \end{array} 
  \right. \quad .
\end{equation}
Thus, the protocol of random movements can be defined by giving
the  p.d.f.'s for the new values of the system variables. This seems a
natural implementation for continuous systems.

The next step consist in estimating $N_{up}$ and $N_{dn}$ for a
given configuration with energy $E_{old}$.
The problem arises because there is a continuum non-numerable set of
configurations that can be reached.
However, it has to be noted that each one of these new configurations
has a well-defined energy value $E_{new}$ and that exactly one of them
is chosen at random.
Therefore, we can define $E_{new}$ and the energy change $\Delta E
\equiv E_{new} - E_{old}$ as random variables. 
The p.d.f. $f_{\Delta E}(\Delta E)$ is
defined by the probability of obtaining an energy change between
$\Delta E$ and $\Delta E+d\Delta E$.
We propose to redefine $N_{up}$ and $N_{dn}$ as
\begin{eqnarray}
  \label{NupNdn}
  N_{up} \equiv f_{\Delta E}(\Delta E_{fix}) & ; &
  N_{dn} \equiv f_{\Delta E}(- \Delta E_{fix})
\end{eqnarray}
From this point, the method proceeds the same as in the
discrete case. 
Eqs. (\ref{NupNdn}) extend the
BHMC method to systems with continuous degrees of freedom.
Therefore, the problem reduces to
finding the function $f_{\Delta E}$ for a given configuration $X_{old}$.

\subsection{Finding $f_{\Delta E}$}

The procedure of finding $f_{\Delta E}$ is
straightforward. 
First, the protocol of random movements is defined by giving the
p.d.f.'s for the new values of the system variables, as in
Sec. \ref{BHMCcontinuo}.
Then, the energy $E_{new}$, is expressed in terms of such new values.
Next, its p.d.f. $f_{E_{new}}$ is determined by using the
usual rules of finding the p.d.f. for a function of random variables
\cite{random}.
Finally $f_{\Delta E}(\Delta E)$ is found, according to these
rules, replacing $E$ in $f_{E_{new}}(E)$ by
$ \Delta E + E_{old}$.

To continue with our example, we analyze first what happens with the
turn of one specific spin.
Suppose that we choose to turn spin $i$. Its bonds
are the only ones that change the energy value. We define
$\varepsilon_i$ as sum of the energies of these bonds, i.e. 
\begin{equation}
  \label{Ei1}
  \varepsilon_i \equiv - \sum_{j} \cos {(\theta_i - \theta_j)} \quad .
\end{equation}
where the summations are taken over all nearest neighbors $j$ of the
site $i$.
When $\theta_i$ changes from its previous value
${\theta_i}_{old}$ to the new value ${\theta_i}_{new}$,
$\varepsilon_i$ changes from ${\varepsilon_i}_{old}$ to
${\varepsilon_i}_{new}$.
It is clear that $\Delta E = {\varepsilon_i}_{new}
-{\varepsilon_i}_{old}$ and that ${\varepsilon_i}_{new}$ is a 
function of the random variable ${\theta_i}_{new}$. 
Therefore, ${\varepsilon_i}_{old}$ and ${\varepsilon_i}_{new}$ can be
used instead of $E_{new}$ and $E_{old}$ for the general treatment
described above. 

In order to express ${\varepsilon_i}_{new}$ in terms of
${\theta_i}_{new}$, Eq. (\ref{Ei1}) can be rewritten as
\begin{equation}
  \label{Ei2}
  {\varepsilon_i}_{new} = A_i \cos {({\theta_i}_{new} -
    \delta \theta_i)} \quad, 
\end{equation}
where
\begin{equation}
  \label{deltatheta}
  \delta \theta_i = \arctan {\sum_{j} \sin \theta_j \over \sum_{j}
    \cos \theta_j} + \pi
\end{equation}
and $A_i$ is defined by Eq. (\ref{A}). 

If ${\theta_i}_{new}$ is a random variable with p.d.f. given by
Eq. (\ref{fThetaNew}),
the usual rules of finding the p.d.f. for a function of
random variables can be used to show that ${\varepsilon_i}_{new}$ is a
random variable with p.d.f. given by \cite{random}
 \begin{equation}
 \label{fEi}
  f_{{\varepsilon_i}_{new}}(\varepsilon)= \left\{ 
  \begin{array}
     {ccc}  
     {1 \over \pi} {1 \over \sqrt{A_i^2-\varepsilon^2}} & : &
     |\varepsilon_i|<A_i
     \\ 0 & : & otherwise
  \end{array} \right. \quad.
\end{equation}

Finally, we replace $\varepsilon$ in Eq. (\ref{fEi}) by
$\Delta E + {\varepsilon_i}_{old}$, to obtain the p.d.f. $f_{
  \Delta E}^i (\Delta E)$ of $\Delta E$ corresponding to a turn
of the spin $i$, i. e. 
\begin{equation}
  \label{fDeltaEi}
 f_{\Delta E}^i (\Delta E)= \left\{ \begin{array} {ccc} 
 {1 \over \pi} {1 \over \sqrt{A_i^2-({\varepsilon_i}_{old}+\Delta E)^2}} 
 & : &|\Delta E+{\varepsilon_i}_{old}|<A_i
 \\ 0 & : &otherwise
\end{array} \right. \quad .
\end{equation}

\begin{figure}[hbt!]
  \begin{center}
    \includegraphics[width=7.5cm,bb=0 0 512 433,angle=0]{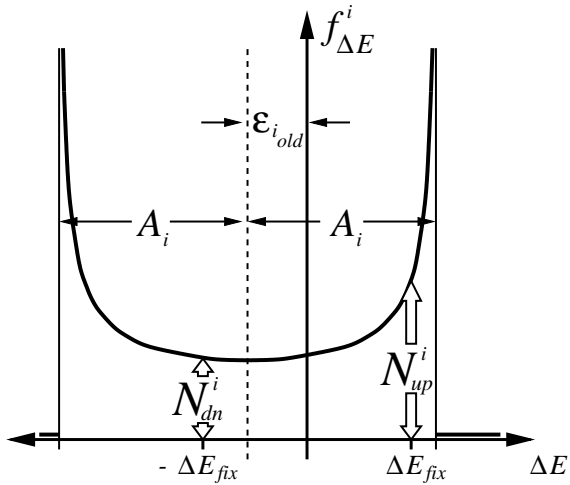}
    \caption{\footnotesize Probability Density function $f_{\Delta E}^i
      (\Delta E)$ of the energy change $\Delta E$ obtained by turning
      spin $i$, according to Eq. (\ref{fDeltaE}). The function
      diverges at points $\Delta E=\pm A_i-{\varepsilon_i}_{old}$, takes
      its minimum value between these points at $\Delta
      E=-{\varepsilon_i}_{old}$  and
      equals zero outside this range. The contributions
      $N_{up}^i$ and $N_{dn}^i$ of turning the spin $i$ are shown.}
    \label{fDE}
  \end{center}
\end{figure}

Since all spins are equally probable,
the probability of obtaining an energy change $\Delta E$ between $\Delta
E$ and $\Delta E+d \Delta E$ is equal to the sum of
$f_{\Delta E}^i (\Delta E)$ over all spins divided by $N$, the
number of spins, i.e. 
\begin{equation}
  \label{fDeltaE}
 f_{\Delta E} (\Delta E) 
 = {1 \over N}\sum_i f_{\Delta E}^i (\Delta E) \quad .
\end{equation}
$N_{up}$ and $N_{dn}$ for a given configuration are obtained finally
using Eqs. (\ref{NupNdn}) as
\begin{eqnarray}
  \label{XYNupNdn}
  N_{up} = {1 \over N}\sum_i N_{up}^i &;&
  N_{dn} = {1 \over N}\sum_i N_{dn}^i
\end{eqnarray}
where we have defined
\begin{eqnarray}
  \label{NupNdni}
  N_{up}^i \equiv f_{\Delta E}^i (\Delta E_{fix}) &;&
  N_{dn}^i \equiv f_{\Delta E}^i (- \Delta E_{fix})
\end{eqnarray}
When Eqs. (\ref{XYNupNdn}) are replaced in Eq.(\ref{diff}), the factor
${1 / N}$ in
Eq. (\ref{fDeltaE}) as well as the factor ${1 / \pi}$ in Eq. 
(\ref{fDeltaEi}) will eventually cancel out.

The p.d.f. $f_{\Delta E}^i (\Delta E)$ is shown in Fig. \ref{fDE}. The
function diverges at points $\Delta E=\pm A_i-{\varepsilon_i}_{old}$ and
takes its minimum value between these two points at $\Delta
E=-\varepsilon_{old}$ with $f_{\Delta E}^i (-{\varepsilon_i}_{old})=(\pi
A_i)^{-1}$. Outside this range the function equals zero. The values
$N_{up}^i$ and $N_{dn}^i$ are also shown.

Summarizing, Eqs. (\ref{XYNupNdn}) give
us the way to calculate  $N_{up}$ and $N_{dn}$ for the XY-model and,
therefore, a way to apply the BHMC method to it.
The same strategy can be used to apply the method to other
continuous systems.

\section{How to take the samples}
  \label{Sampling}

A second idea behind the BHMC method is to take the samples in such a way
that they distribute almost homogeneous over the whole energy range of
interest. There are many different strategies that can be used for
this purpose. We have used two of them, namely, a
micro-canonical sampling process that maintains the system inside a narrow
energy window (BHMC-$\mu$C) \cite{bhmc3}, and a random walk on the
energy axis using Metropolis \cite{Metropolis} steps (BHMC-M) \cite{bhmc1}.

The BHMC-$\mu$C sampling strategy consist in taking a small window
centered at each 
energy of interest, starting from a configuration inside the
window and performing random movements never surpassing the limits of the
window.
It can be implemented, for instance, by taking at random a new configuration
and accepting it only if it falls
inside the window, or by using the Creutz energy bag method
\cite{DaemCreutz}.
According to Eq. (\ref{beta}), two micro-canonical simulations are
required to calculate $\beta (E)$ at one energy value. However,
the data of one simulation can be used for two energy values if their
difference equals $\Delta E_{fix}$.
For the XY-model, we take one spin 
and generate at random a new angle $\theta \in [-\pi,\pi]$.
If the total
energy of the system falls inside the window, the change is
accepted, otherwise it is rejected. 
It is clear that this sampling process maintains a detailed balance
condition with equal probabilities for the old an new configurations
and, therefore, it samples all states inside the window with the same
probability.
A whole lattice sweep is performed
by repeating this procedure for all spins of the system, and a new
configuration is taken after every fixed number of lattice sweeps. 

The BHMC-M sampling strategy consist first in performing
a Markovian process with symmetric probability
distributions for increasing and decreasing the energy, i.e. a non
biased random walk on the energy axis. Second,
since the canonical distribution is almost symmetric on
the energy axis around its mean energy value $<E>_T$, 
one way to do it (but not the only one) consists of taking the last
configuration of the sample and performing on it a fixed number of
Metropolis \cite{Metropolis} steps at such a temperature that its
canonical distribution appears centered on the energy value $E$. The
resulting configuration is taken as the next sample.
In order to find this temperature $T(E)$ Eq. (\ref{beta}) can be used 
by remembering that $\beta (E) \equiv {1/T(E)}$.
Furthermore, Eq. (\ref{diff}) can be used to estimate $\beta(E)$ and
$T(E)$ in more accurate ways. 
In the present paper we use the centered difference estimator
\begin{eqnarray}
 \label{beta2}
 \beta(E) &\simeq&
 {\ln g(E+\Delta E_{fix})-
  \ln g(E-\Delta E_{fix}) \over 
  2\Delta E_{fix}} \nonumber \\ &\simeq&
 {1 \over 2 \Delta E_{fix}} \ln {<N_{up}(E)>
   <N_{up}(E-\Delta E_{fix})>
   \over 
   <N_{dn}(E)><N_{dn}(E+\Delta E_{fix})>}.
\end{eqnarray}
These values are estimated from the data accumulated in the
histograms $N_{up}(E)$, $N_{dn}(E)$ and $V(E)$.
It can happens that the system reaches an energy value $E$ such that
some of the neighboring energies have yet not been sampled and,
therefore, $T(E)$ cannot be estimated. In such a case, the
Metropolis steps are performed at the same temperature used to
produce the last configuration. 

\section{Implementation}
  \label{SecImplementa}

In order to test our strategy numerically we performed computations
for the three-dimensional XY-model on a $10 \times 10 \times 10$ cubic
lattice with periodic boundary conditions using the BHMC-M, the
BHMC-$\mu$C, the HMC and the Metropolis method.

For the Metropolis method the spin system was 
initialized at each temperature value from a random configuration
performing $N_{ini}=500$ entire Metropolis lattice sweeps  before
sampling, and then $N_s=1500$ samples were taken separated by $N_i=10$
lattice sweeps among them in order to decrease their correlations.
Here and in the
following the correlation between successive samples was determined by
looking at the correlation of their magnetization values.
For the Metropolis method we obtained a correlation between successive
samples of $56.5\%$ at $T_{\rm o}=2.159$, i.e. at the pseudo critical
temperature we have observed for this finite system.
Some
additional Metropolis simulations were performed as reference data for
the comparison between the BHMC-$\mu$C and the HMC methods. 
These simulations were performed in the same way as before, but taking
$N_s=10000$ samples at each temperature value.

For the BHMC-$\mu$C method we divided the whole energy range of
positive temperatures, i.e. energies per bond between $-1.0$ and
$0.0$, in adjacent windows of equal size and chose $\Delta E_{fix}$
to be equal to this bin size. So $\Delta E_{fix}$ is equal to the
difference of the energy values at which consecutive micro-canonical
simulations are performed and, therefore, the data of each simulation
can be used twice (see Sec. \ref{Sampling}). 
To acquire the samples the spin system was 
initialized for the lowest-energy window from a random configuration
by turning some spins at random until the window is reached and
performing $N_{ini}$ micro-canonical lattice sweeps before
sampling. Then, we took $N_{s}$
 samples separated by $N_i$ lattice sweep among them. After
that, we move to the next window by turning some spins at random
starting from the previous sample. Then, $N_{ini}$
micro-canonical lattice sweeps were performed before sampling, and so
on.

To compare with the Metropolis simulations, we divided the whole
energy range of positive temperatures in $1225$ adjacent windows,
i.e. $\Delta E_{fix} \simeq 2.45$.
To handle the critical slowing down in the
relaxation time for the micro-canonical simulations,
larger values of $N_{ini}$ have
been taken in the vicinity of the critical point, i.e.
$N_{ini}=250$ for energies per bond between $-0.400$ and $-0.320$
and $N_{ini}=70$ for energies outside this range.
To obtain the same correlation between samples as from Metropolis
simulations we used $N_i=3$ in the energy range $(-0.429,-0.282)$ and
$N_i=2$ outside this range. The maximal correlation in the vicinity of
the critical point so obtained was $56.4 \%$.
We took $N_{s}=50$ for the whole energy range.

To compare with the HMC method, the whole energy range of
positive temperatures was divided in $500$ adjacent windows,
i.e. $\Delta E_{fix}=6.0$. 
To handle the critical slowing down, we took
$N_{ini}=200$ for energies per bond between $-0.400$ and $-0.320$, and
$N_{ini}=80$ for energies outside this range.
$N_i=2$ was enough to obtain lower correlations in the vicinity of the
critical point than from Metropolis simulations. The maximal
correlation so obtained was $46.3\%$.
In order to obtain the same accuracy than the HMC method on the
temperature range $2.1<T<2.2$ more samples have been taken in the
vicinity of the critical point,
i.e. $N_{s}=380$ for energies per bond between $-0.504$ and $-0.248$
and $N_{s}=80$ for energies outside this range.

For the BHMC-M method the histograms were constructed by dividing the 
whole energy range of
positive temperatures in $1225$ boxes of equal size.  
Ten independent systems that accumulated data in common histograms
were simulated simultaneously, in order to estimate $T(E)$ more
accurately. 
These systems were initialized from random configurations thermalized
by ${N_{ini}=500}$ Metropolis lattice sweeps at the critical temperature
  ${T_{\rm c}=2.20196}$ \cite{XY_Adler}. 
We took a total number of ${N_s=10 \times 10000}$ samples performing 
$N_i=1$ lattice sweeps between them.
In order to avoid wasting samples at very low temperatures,
we restricted our systems to energies between $-0.90$ and
$0.0$ per bond; spin systems going out of this range were reinitialized
by copying the values of an additional spin system, which acts as a
replacement. This additional system was initialized in the same way
as the other ones, and then twenty entire Metropolis lattice
sweeps at $T_{\rm c}$ were applied on it after each time it was
taken as replacement.

The HMC method was implemented by dividing the whole energy range of
positive temperatures in $1225$ and $500$  boxes of equal size, and
accumulating the data of a Metropolis simulation at $T_{\rm o}=2.159$
performed with $N_{ini}=500$,$N_i=10$ and $N_s=10000$.
  
For all methods, we stored $\cos (\theta_i)$ and $\sin (\theta_i)$ in
tables instead of the angles $\theta_i$ in order to accelerate the
calculations. For the BHMC-$\mu$C and BHMC-M methods, we stored in
addition $A_i^2$ and $\varepsilon_i$ for each spin for the same
reason. The functions $\cos(x)$, $\sin(x)$, $\sqrt x$, $1/\sqrt x$,
$\exp(x)$, were stored in tables for all methods.

The mean values and error bars for each method were made by
repeating eight times the whole simulation with 
different initial seeds of the random number generator. We
used the ``minimal'' Park and Miller random number generator combined
with a Marsaglia shift sequence from Ref. \cite{NumRCPS_F90}. 

\section{Results}
 \label{results}

\begin{figure}[htb!]
  \begin{center}
    \includegraphics[width=7.5cm,bb=45 45 535 445,angle=0]{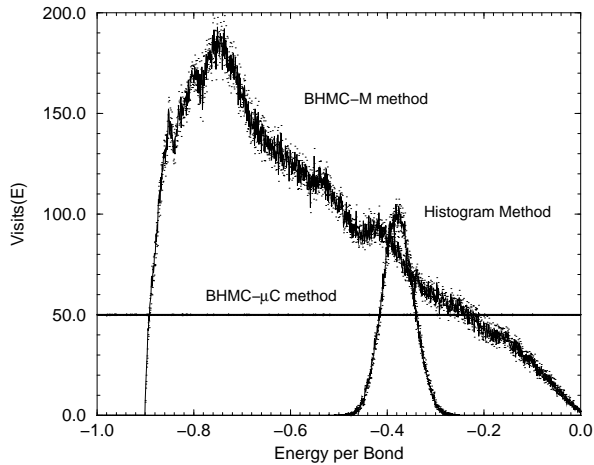}
    \caption{\footnotesize Number of visits as a function of energy obtained
      from the BHMC-$\mu$C method with $50$ samples per window, the
      BHMC-M method with $10 \times 10000$ samples, and the canonical 
      histogram method with $10000$ samples for
      the $10 \times 10 \times 10$ XY-model on a cubic lattice.
      All three calculations were performed by using $1225$ divisions on
      the energy axis.
      The points over and below each curve correspond to
      the upper and lower limits of the error bars. On this energy scale
      the mean energy at the critical temperature observed
      corresponds to $E=-0.377$.}
    \label{Visits}
  \end{center}
\end{figure}

\begin{figure}[hp!]
  \parbox{\textwidth}{
    \begin{center}
    \includegraphics[width=7.5cm,bb=45 45 535 445,angle=0]{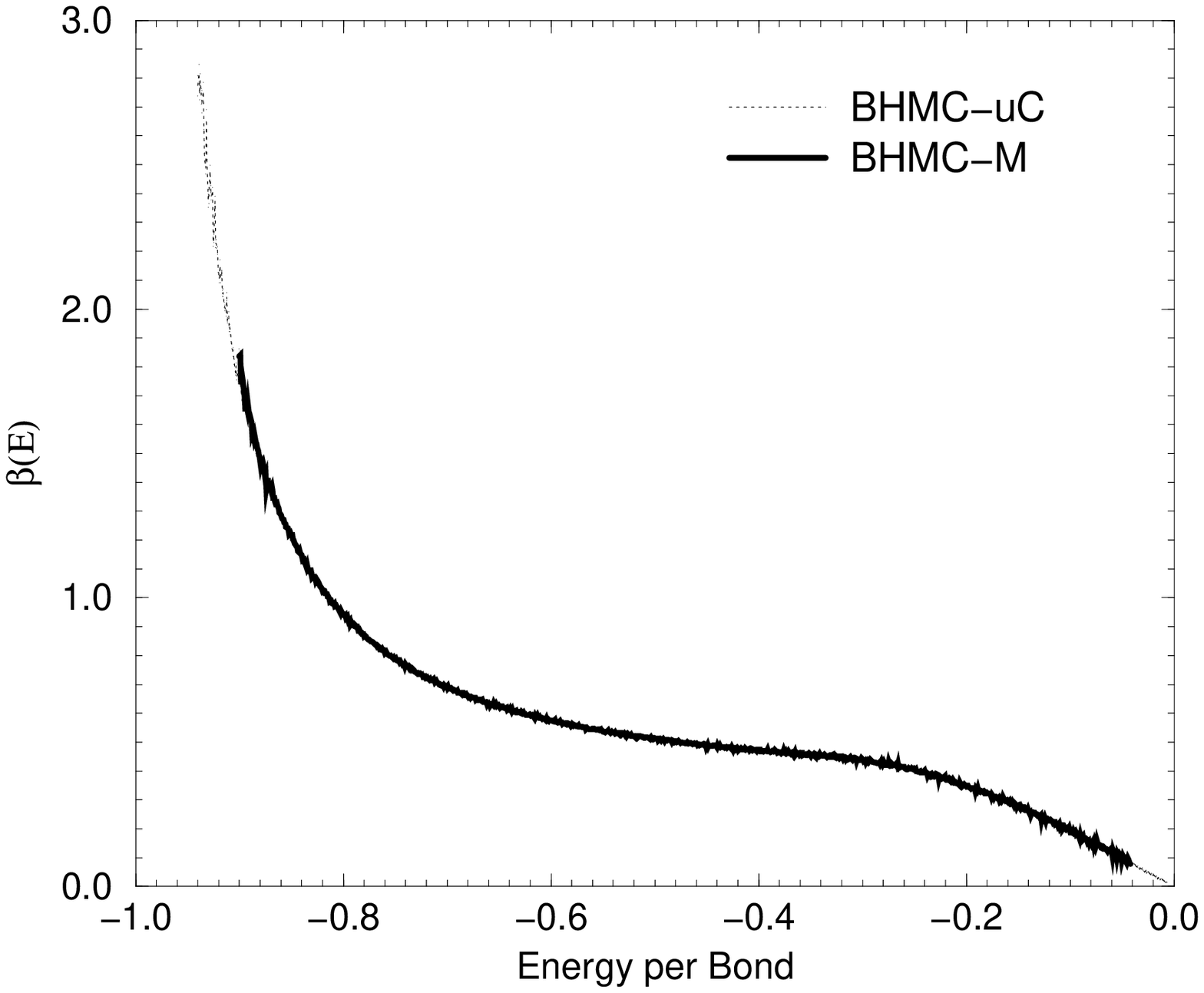}
    \caption{\footnotesize $\beta \equiv 1/T$ as a function of the energy
    for the SC ${10 \times 10 \times 10}$ XY-model obtained from
    the BHMC-$\mu$C method (dotted line)
    and the BHMC-M method (solid line).
    Both calculations were performed with ${\Delta E_{fix} \simeq 2.45}$.
    The error bars are displayed as in Fig. \ref{Visits}.
    These errors are so small, that
    their points are almost indistinguishable from the curves.}
    \label{FigBeta}
  \end{center}}
  \parbox{\textwidth}{
    \begin{center}
    \includegraphics[width=7.5cm,bb=45 45 535 445,angle=0]{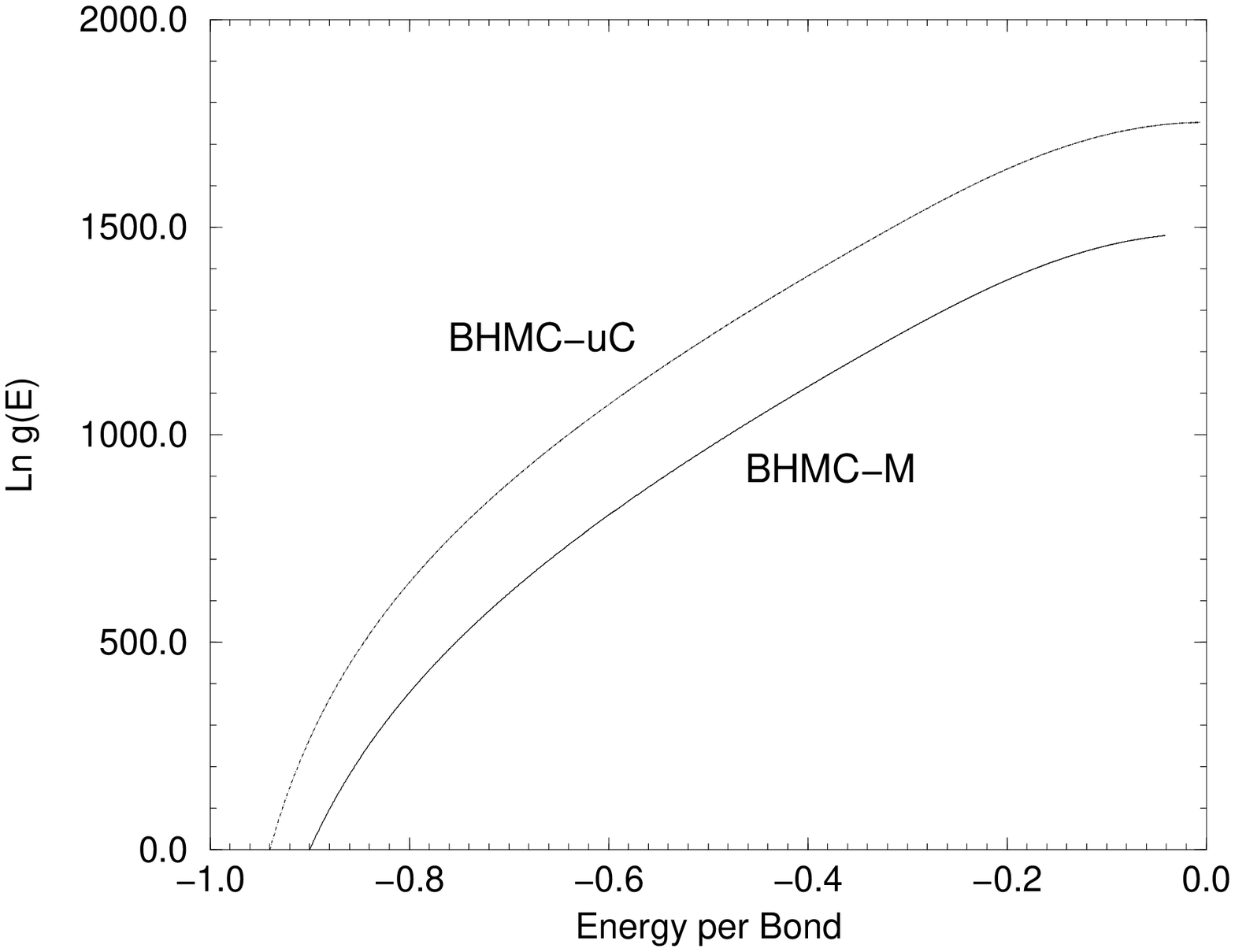}
    \caption{\footnotesize Degeneracy $g(E)$ for the SC ${10 \times 10 \times
    10}$ XY-model obtained from the BHMC-$\mu$C 
    and the BHMC-M methods.
    Both calculations were performed with ${\Delta E_{fix} \simeq 2.45}$.
    The error bars are displayed as in Fig. \ref{Visits}.
    The curves differ only in a constant value,
    namely $\ln (g(-0.90)/g(-0.94))$.}
    \label{FigLngE}
  \end{center}}
\end{figure}

\begin{figure}[hp!]
  \parbox{\textwidth}{
    \begin{center}
      \includegraphics[width=7.5cm,bb=45 45 535 445,angle=0]{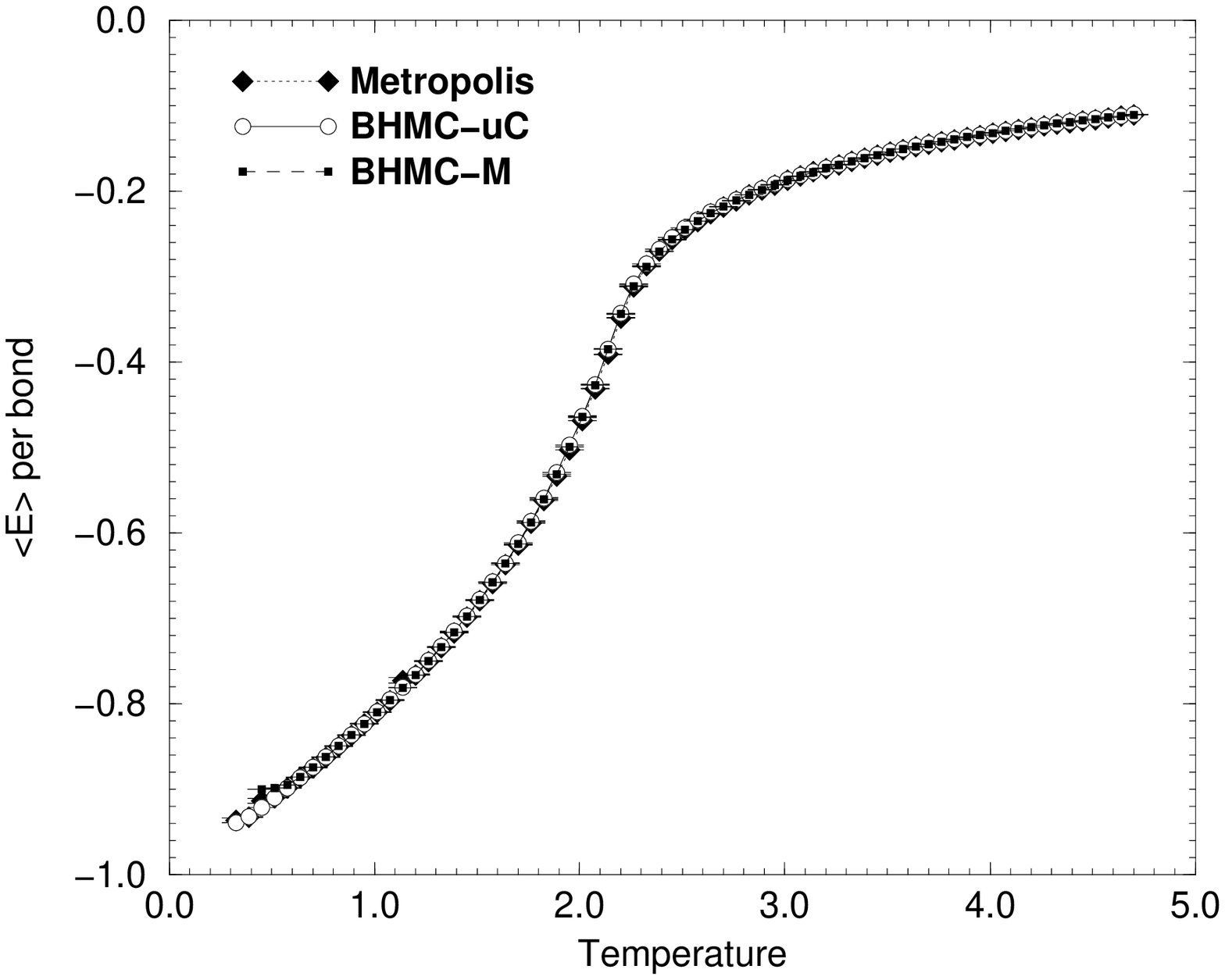}
      \caption{\footnotesize Averaged energy for the SC ${10 \times 10 \times
          10}$ XY-model 
        obtained from the BHMC-$\mu$C method (circles), the BHMC-M
        method (small squares), and Metropolis simulations (diamonds).
        Both BHMC calculations were performed with ${\Delta E_{fix} \simeq 2.45}$.
        The values obtained from the three procedures
        are almost indistinguishable.}
      \label{FigE}
    \end{center}
    }
  \parbox{\textwidth}{
    \begin{center}
      \includegraphics[width=7.5cm,bb=45 45 535 445,angle=0]{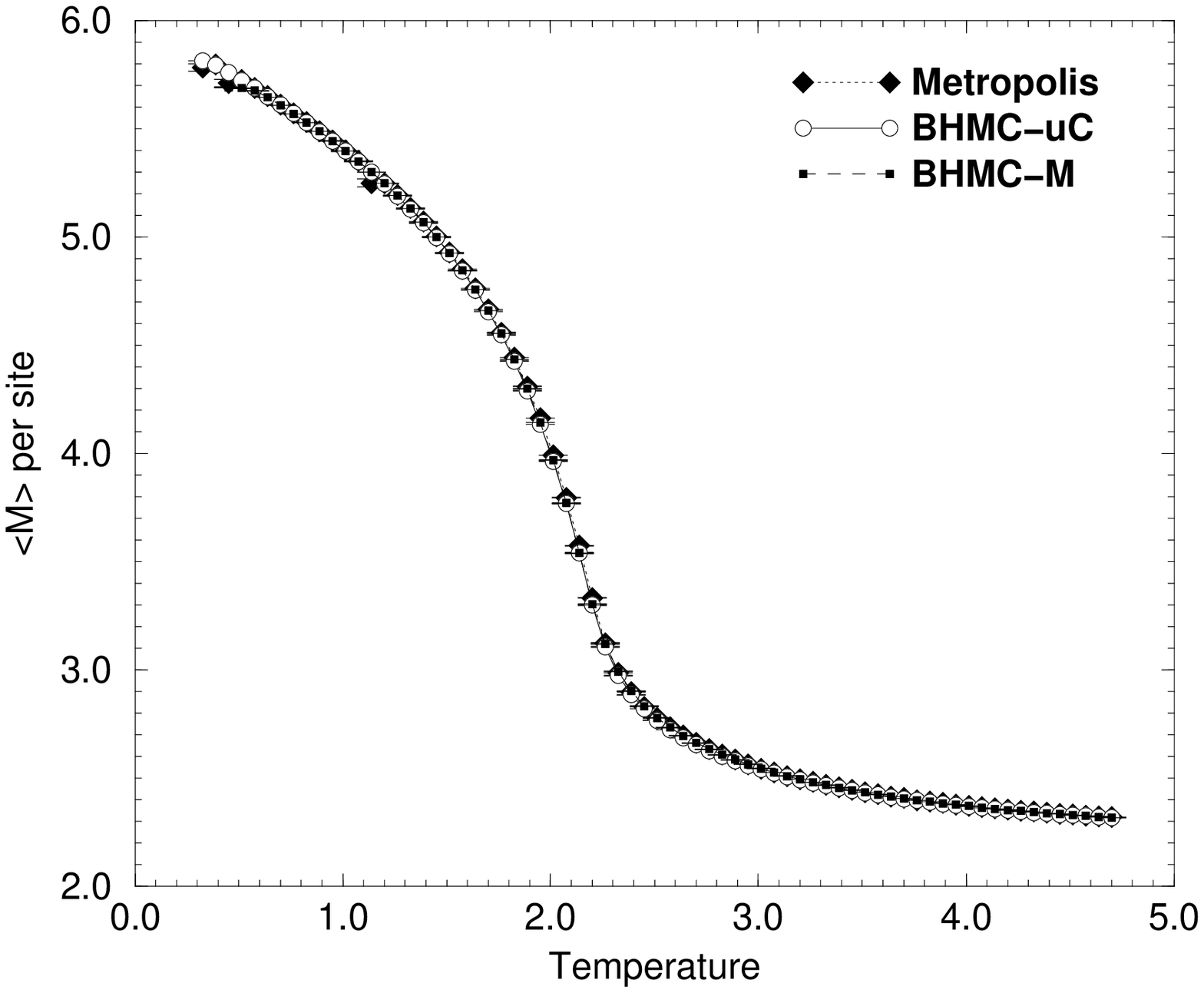}
      \caption{\footnotesize Averaged magnetization for the  SC ${10 \times 10
          \times 10}$ XY-model 
        obtained from the BHMC-$\mu$C method 
        (circles), the BHMC-M method 
        (small squares) and Metropolis simulations (diamonds).
        Both BHMC calculations were performed with ${\Delta E_{fix} \simeq 2.45}$. 
        As in Fig. \ref{FigE}, the three curves are almost indistinguishable.}
      \label{FigM}
    \end{center}
    }
  \end{figure}

\begin{figure}[hp!]
  \parbox{\textwidth}{
  \begin{center}
    \includegraphics[height=5.7cm,width=7.0cm,bb=45 45 535 445,angle=0]{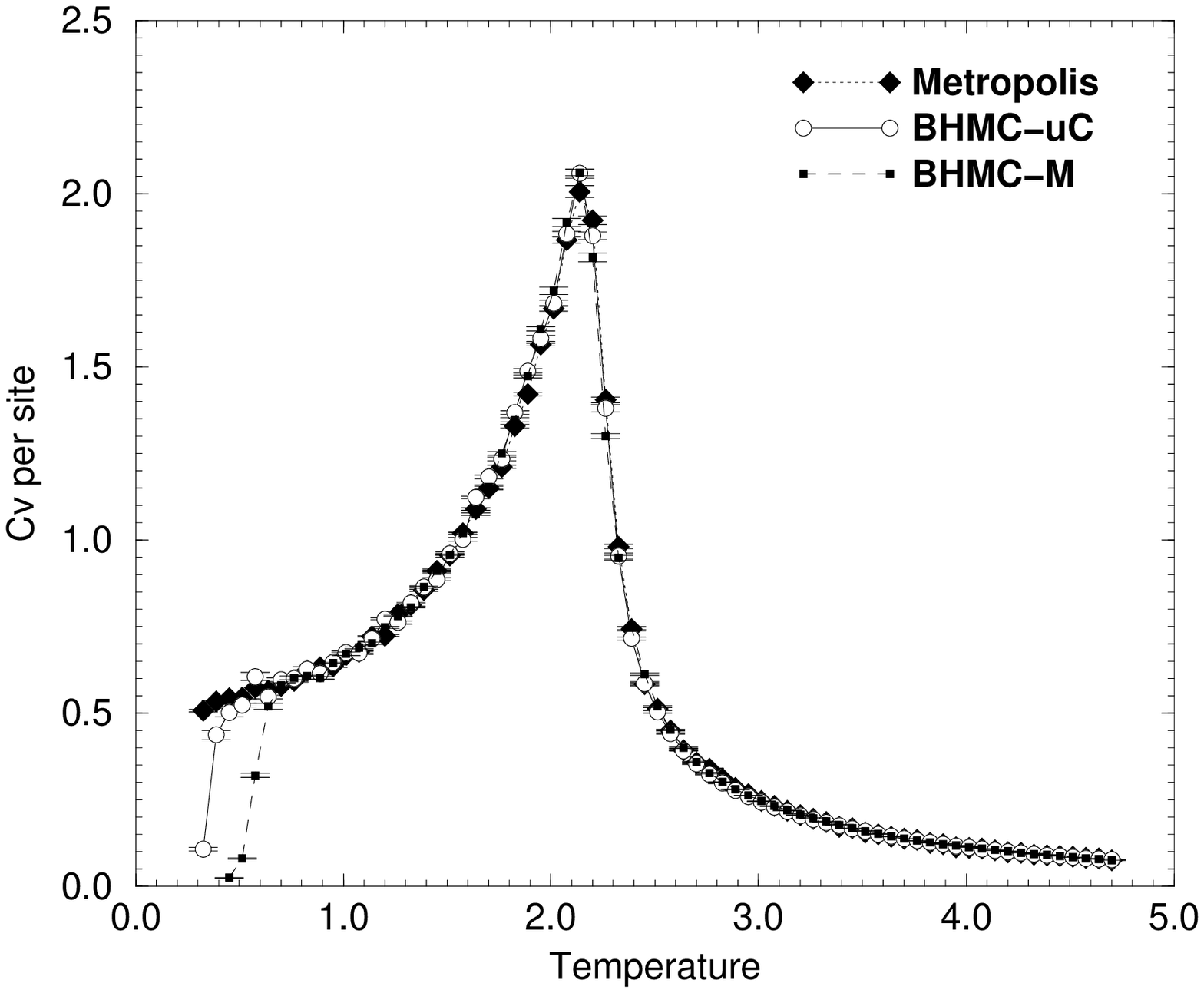}
    \caption{\footnotesize Specific heat for the  SC ${10 \times 10 \times 10}$
    XY-model 
    obtained from the BHMC-$\mu$C method (circles), 
    the BHMC-M method (small squares)
    and Metropolis simulations (diamonds).
    Both BHMC calculations were performed with ${\Delta E_{fix} \simeq 2.45}$.
    Within the error bars, the curves fit very well onto each
    other over the temperature range $0.7<T<4.7$.
    The largest differences appear near $T_{\rm c}$, as would be
    expected. This figure gives an approximate critical temperature
    of $T_{\rm c}=2.16$ for the finite size $10 \times 10 \times 10$
    system.}
    \label{FigCv}
  \end{center}}
  \parbox{\textwidth}{
  \begin{center}
    \includegraphics[height=5.7cm,width=7.0cm,bb=45 45 535 445,angle=0]{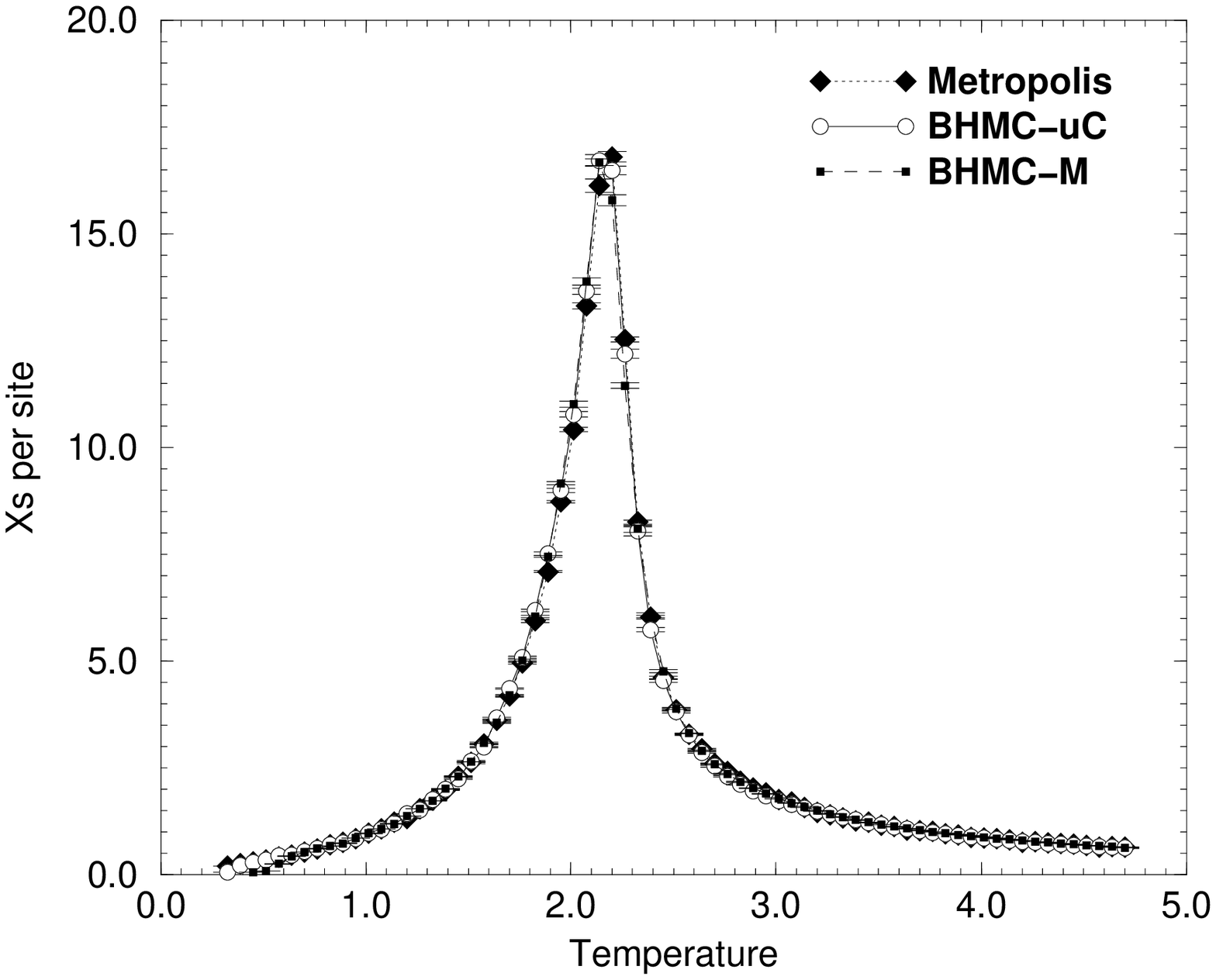}
    \caption{\footnotesize Magnetic susceptibility for the SC ${10 \times 10
    \times 10}$ XY-model 
    obtained from the BHMC-$\mu$C method (circles),
    the BHMC-M method (small squares)
    and Metropolis simulations (diamonds).
    Both BHMC calculations were performed with ${\Delta E_{fix} \simeq 2.45}$.
    As in Fig. \ref{FigCv},
    the values obtained from the two methods fit very well onto each
    other, showing the largest differences at temperatures near
    $T_{\rm c}=2.16$.}
    \label{FigXs}
  \end{center}}
\end{figure}

\begin{figure}[hp!]
  \parbox{\textwidth}{
    \begin{center}
    \includegraphics[height=5.7cm,width=7.0cm,bb=45 45 535 445,angle=0]{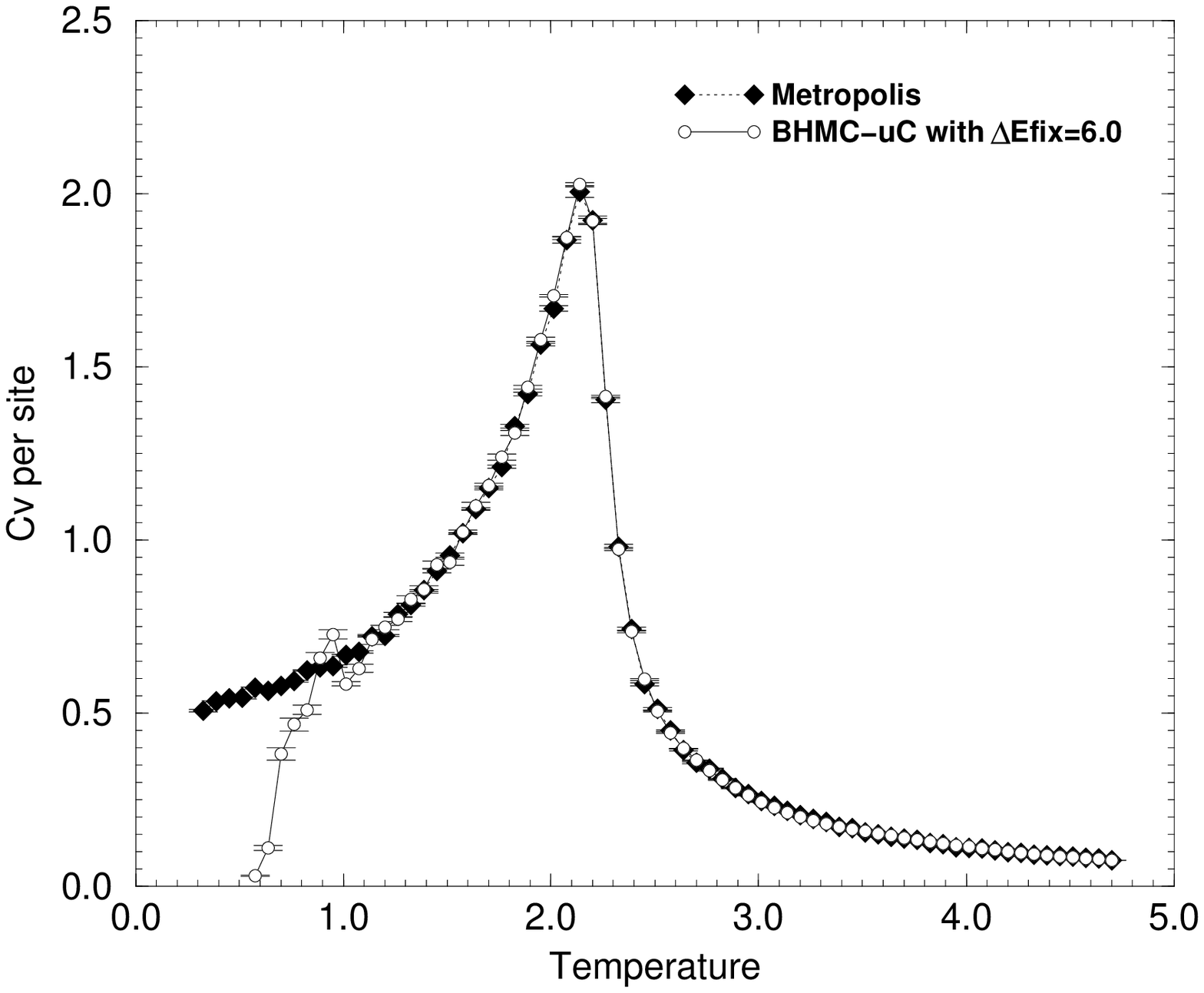}
    \caption{\footnotesize Specific heat for the  SC ${10 \times 10 \times
        10}$ XY-model obtained from 
    the BHMC-$\mu$C method with ${\Delta E_{fix}=6.0}$ (circles)
    and Metropolis simulations (filled diamonds). 
    The curves fit very well onto each other over the temperature range
    $1.2<T<4.7$.}
    \label{FigHMC1}
  \end{center}}
  \parbox{\textwidth}{
    \begin{center}
    \includegraphics[height=5.7cm,width=7.0cm,bb=45 45 535 445,angle=0]{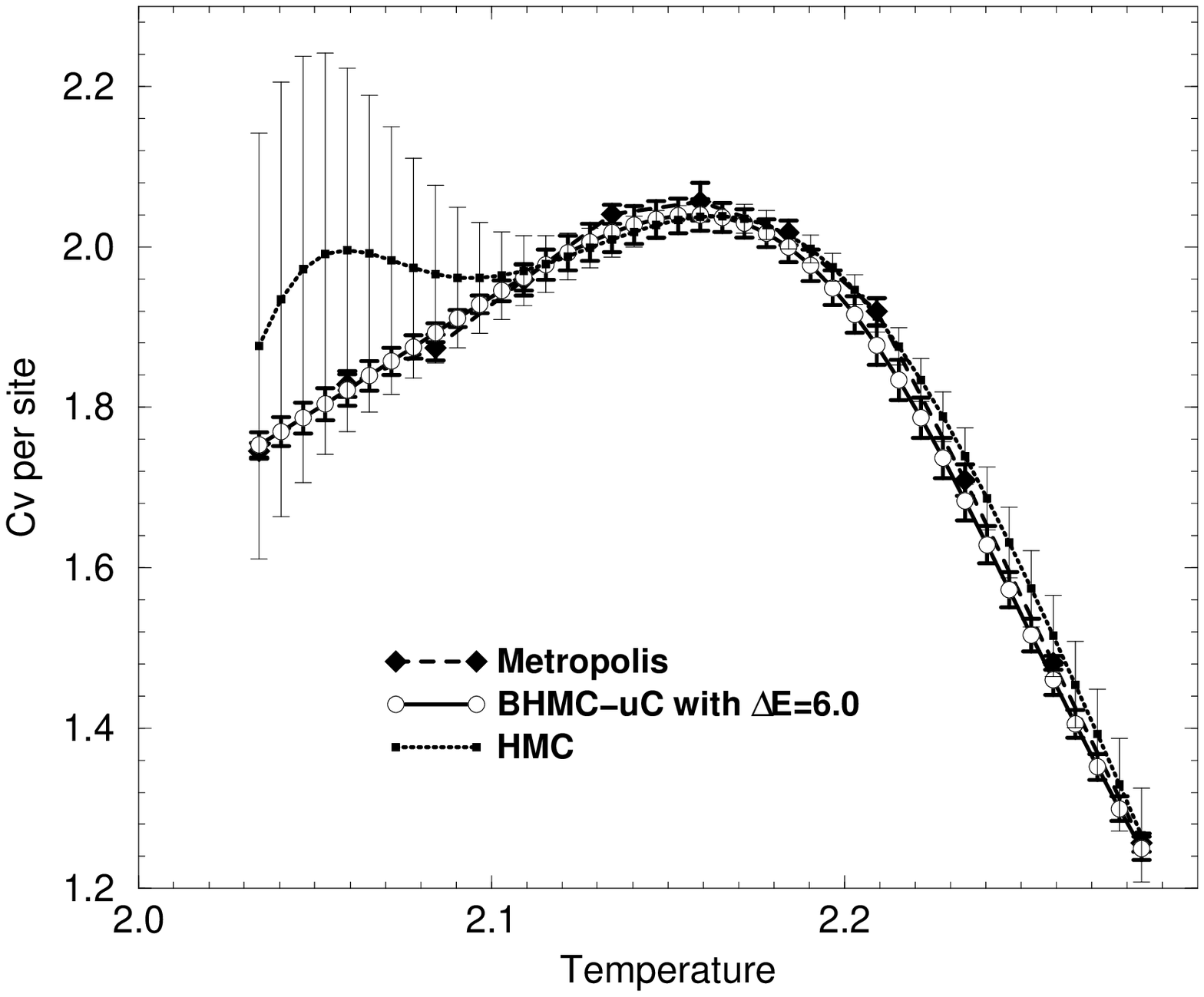}
    \caption{\footnotesize Specific heat for the  SC ${10 \times 10 \times
        10}$ XY-model obtained from 
    the BHMC-$\mu$C method with ${\Delta E_{fix}=6.0}$ (circles),
    the HMC method with ${\Delta E=6.0}$ (small squares)
    and Metropolis simulations with $10000$ samples (filled diamonds). 
    all methods coincide in the temperature range $2.1<T<2.2$, but the
    HMC method deviates for temperatures outside this range.
    Instead,
    the BHMC-$\mu$C method remains precise up to the range $1.2<T<4.7$
    (see Fig. \ref{FigHMC1}) and using only $2.2$ times more computer
    effort than the HMC method.
    This figure shows error bars of $3.5$ standard
    deviations, i.e. a confidence level of $99\%$ for eight runs.}
    \label{FigHMC2}
  \end{center}}
\end{figure}

The results are shown in
Fig. \ref{Visits}-\ref{FigHMC2}. Figs. \ref{Visits}-\ref{FigXs}
display data obtained by
using $1225$ divisions of the energy range of positive temperatures,
i.e. ${\Delta E_{fix} \simeq 2.45}$ for the BHMC-$\mu$C, the BHMC-M
and the HMC methods.

Fig. \ref{Visits} shows how samples distribute on the energy axis
using the BHMC-$\mu$C, the BHMC-M and the HMC methods.
For the BHMC methods a large energy range 
is sampled, in contrast to the narrow window sampled by
the classical canonical histogram method. In addition, with the BHMC-$\mu$C
it is possible to take exactly the same number of samples at each
energy value. 

Fig. \ref{FigBeta} shows $\beta (E)$ obtained from the BHMC-$\mu$C and
the BHMC-M methods. 
It has to be noted that both curves are almost
indistinguishable, in spite of the completely different sampling
strategies of both methods.

Fig. \ref{FigLngE} shows $\ln g(E)$ obtained from the BHMC-$\mu$C and
the BHMC-M methods.
Since the curve from the BHMC-M method was obtained
adding the differences of Eq. (\ref{diff}) starting from
the point $(E_{\rm o}= -0.90,\ln g(E_{\rm o})=0)$, 
its values $g(E)$ are the ratio of the probability to obtain at
random an energy per site between $E$ and $E+dE$ to the probability to obtain
an energy per site between this value of $E_{\rm o}$ and $E_{\rm o}+dE$, when
all states are equally probable. The same applies to the curve
obtained from the BHMC-$\mu$C method, with $E_{\rm o}=-0.94$. It can
be noted, that both curves are only shifted in a constant value, namely 
$\ln (g(-0.90)/g(-0.94))$.

Figs. \ref{FigE}-\ref{FigXs} compare results from the BHMC methods
with Metropolis simulations.
Figs. \ref{FigE} and \ref{FigM} show the average energy and the
average magnetization obtained with the BHMC-$\mu$C and the BHMC-M
methods compared with the same quantities obtained using Metropolis
simulations.  The agreement between the three Monte Carlo methods is
excellent. The data obtained by the BHMC methods fit so well
with those obtained by Metropolis simulations, that their curves and
their error bars are almost indistinguishable.

In Fig. \ref{FigCv} and \ref{FigXs} we show the heat capacity and the
magnetic susceptibility obtained by using
the BHMC-$\mu$C, the BHMC-M and the Metropolis simulations.
The curves calculated by the BHMC-M method 
fit very well within their error bars with them of the Metropolis
simulations for the temperature
range between $T=4.7$ and $T=0.7$, and deviate for lower
temperatures. The curves calculated by the BHMC-$\mu$C method fit as
well as those from the BHMC-M method in the same temperature range,
begins to oscillate at $T=0.7$ and deviates for a lower
temperature, namely $T=0.45$.
This is due to a lack of good statistics at low
energies. In the case of BHMC-$\mu$C method most of the possible
changes at low energy values increase the system energy and,
therefore, it is difficult
to obtain estimations for $<N_{dn}(E)>$ different from zero at this
energy values. This
difficulty increases when $\Delta E_{fix}$ increases.
In the case of the BHMC-M method we have not sampled the energy range
below $E=-0.9$ at all. It can be observed that this fact only affects
the results at low temperatures.
For all methods, the errors bars become larger around the critical
temperature $T_{\rm c}$, as it would be expected.

For all these curves, we present error bars of $1.0$ standard
deviations. For eight runs, the confidence level of $99\%$ corresponds
to $3.5$ standard deviations.
 
Since all sets of data obtained from the BHMC-$\mu$C method and the
Metropolis simulations present error bars of similar size using the
same correlation between samples, we are
allowed to compare the speeds of the two algorithms.
The Metropolis method took $28.5$ seconds on a Digital Alpha workstation
to make one simulation at one temperature point.  
The BHMC-$\mu$C method took $440$ seconds to perform one calculation on
the whole temperature range using the same machine. 
It means that our method calculates the whole temperature range
$0.7<T<4.7$ using approximately $15.5$ times the computer effort
required by a Metropolis simulation for one point. In addition,
the BHMC-M method took $409$ seconds per run.

Figs. \ref{FigHMC1} and \ref{FigHMC2} compare the heat capacity obtained from
the BHMC-$\mu$C 
and the HMC methods using Metropolis simulations as reference
data. Both figures display data obtained from the BHMC-$\mu$C and the
HMC methods using $500$ divisions of the energy axis, i.e. ${\Delta
  E=6.0}$. 
Fig. \ref{FigHMC1} presents the usual error bars of $1.0$ standard
deviations and Fig. \ref{FigHMC2} presents error bars of  $3.5$ standard
deviations corresponding to a confidence level of 99\% for eight runs,
i. e. the probability of finding the true value outside these
error bars due only to statistical variations is equal to $1\%$.
It can be observed that all methods coincide in the temperature
range between $T=2.1$ and $T=2.2$. This is the temperature range of
validity of the HMC method according to the criteria that the mean
energy value calculated at temperature $T$ should be less than $\sigma_E$  
away of the mean energy value at $T_{\rm o}$, where $T_{\rm o}$ is the
temperature used to generate the samples and $\sigma_E$ is the
standard deviation on the energy of these samples \cite{HMCerror}.
For temperatures out of this range, the HMC method loses its precision
but the BHMC-$\mu$C method is still precise on the whole temperature
range $1.2<T<4.7$.
Since the error bars in the range $2.1<T<2.2$ are similar in size and
the correlations between samples are also similar, we can
compare the speeds of both methods.
The HMC method took $188$ seconds per run on a Digital Alpha
workstation. The BHMC-$\mu$C method took $411$ seconds per run using
the same machine. It means that our method gives precise results for
the temperature range $1.2<T<4.7$, i.e. $35$ times larger than the
range $2.1<T<2.2$ of precise results given by the HMC method, using
only $2.2$ times more CPU time.

\section{Conclusions}
 \label{conclusions}

We have proposed a way of applying the BHMC method to continuous
systems.
We have found results for the three-dimensional SC XY-model 
in  excellent agreement with the HMC method and  Metropolis
simulations. Our method calculates the temperature range $0.7<T<4.7$
using less computer time than the Metropolis simulations for $16$
points, and calculates with precision the temperature range
$1.2<T<4.7$ using only $2.2$ times more computer effort than the HMC
method for the range $2.1<T<2.2$. 

The strategy proposed could also be applied to other thermodynamic
systems with continuous degrees of freedom, for example, the
Heisenberg model and fluid models. Our strategy can be summarized as
follows: 

\begin{itemize}
\item Choose a protocol of random movements in the space of states
     of the system, such that for each allowed movement the
     probability to perform it equals the probability to revert it.
     These movements are only virtual, in the sense that they are
     not executed. They are introduced only to calculate $Nup$,
     $Ndn$ and, therefore, to estimate the density
     of states g(E). The protocol would change the system from an
     $old$ to a $new$ state. Express the protocol giving the
     probability density function (p.d.f.) for the values of the
     system variables at the $new$ state. These new values are,
     therefore, random variables. 

\item Find the p.d.f. $f_{\Delta E}(\Delta E)$,
 i.e. the probability of obtaining an energy change between
 $\Delta E$ and $\Delta E+d\Delta E$ using the chosen
 protocol. This can be done as follows.
 \begin{enumerate}
   \item Express the energy of the $new$ state $E_{new}$ as a
     function of the new values of the system variables. $E_{new}$ is,
     therefore, a random variable. 
   \item Use the usual rules of finding the p.d.f. for a function of
     random variables to obtain $f_{E_{new}}(E)$, i.e. the
     p.d.f. of $E_{new}$.
   \item Replace $E$ in $f_{E_{new}}(E)$ by
     $ \Delta E + E_{old}$ to obtain, $f_{\Delta E}(\Delta
     E)$. Here $E_{old}$ denotes the energy of the old state. 
 \end{enumerate}
\item Take $N_{up}$ ($N_{dn}$) for the $old$ state as
  proportional to $f_{\Delta E}(\Delta E_{fix})$ \newline 
  ($f_{\Delta E}(- \Delta E_{fix})$), and
  continue with the BHMC method as usual. $\Delta E_{fix}$
  denotes the fixed energy interval of the BHMC method.
\end{itemize}

The BHMC method calculates the degeneracy $g(E)$ either by adding terms of
the form $\ln g(E+\Delta E_{fix})-\ln g(E)$ or by integrating $\beta
(E)$, starting in both cases from a given initial point $(E_{\rm
  o},\ln g(E_{\rm o}))$. 
If $\ln g(E_{\rm o})=0$ is taken, $g(E)$ can be reinterpreted as the
ratio of the probability to obtain at random an energy between $E$ and
$E+dE$ to the probability to obtain an energy between $E_{\rm o}$ and
$E_{\rm o}+dE$.
This interpretation of the values $g(E)$ obtained from the BHMC method
is valid both for discrete and continuous systems.

Many different strategies can be used to take the samples in the BHMC
method.
In the present paper we used two of them, namely, a
micro-canonical sampling process that maintains the system inside a narrow
energy window (BHMC-$\mu$C) \cite{bhmc3}, and a random walk on the
energy axis using Metropolis \cite{Metropolis} steps (BHMC-M)
\cite{bhmc1}. Between these two methods, we prefer the first one,
because of the following reasons.
First, it resembles better the micro-canonical character of the
averages $<N_{up}(E)>$ and $<N_{dn}(E)>$.
Second, it offers a fine control on the correlations between samples
and the relaxation time at each energy value.
Third, two micro-canonical simulations are enough to estimate $\beta
(E)$ at any desired energy and this fact is independent of the system
size. Therefore, it seems that the computer time will grow
linearly with  the
system size but, of course, the possibility of using data of
one simulation for two energy points discussed in Sec. \ref{Sampling}
will be lost. This will be studied in a future paper.
Fourth, many other kinds of micro-canonical simulations can be employed
to take the samples, different from the accepting-rejecting process used
in this paper but maintaining also a detailed balance condition with equal
probabilities for all configurations inside a window. For instance,
the new configurations can be produced to be always inside the
window, and this will decrease relaxation and correlation times
\cite{DaelCare,Lee}. Another option would be to use Creutz's
demons to take the samples\cite{DaemCreutz}.

Other possible sampling strategy is the Entropic Sampling of Ref.
\cite{EntropicSample1,EntropicSample2}. To use it in conjunction with
the BHMC method 
is a very interesting idea, because the BHMC 
method gives direct estimations of $\ln g(E+\Delta E_{fix})-\ln
g(E)$ and, therefore, no previous estimation of the entropy is
needed to perform the sampling process. Something similar has been
employed in the original work of Ref. \cite{bhmc1}, but estimating
these differences only from the last sample.

In conclusion, the strategy proposed here in order to apply the BHMC
method to systems with continuous degrees of freedom gives excellent
results for the three dimensional XY-model and seems feasible of being
applied to other thermodynamic systems.
It has to be noted, that the BHMC method seems a very precise method to
find $\beta (E)$ and $\ln g(E)$ for every thermodynamic system and
it seems also extensible to the calculation of $g(E)$ for
non-thermodynamic systems.

\section{Acknowledgments}

We thank P.M.C. de Oliveira for helpful discussions. One of the
authors (Mu\~noz) thanks Fernec Kun.
The work of one of the authors (Mu\~noz) is supported by the Deutscher
Akademischer Austauschdienst (DAAD) through scholarship A/96/0390.

\end{document}